\documentclass[aps,pra,twocolumn,showpacs,preprintnumbers,amsmath,amssymb,superscriptaddress,longbibliography]{revtex4-1}

\usepackage{mathrsfs}
\usepackage{amsfonts}
\usepackage{amsmath}
\usepackage{amssymb}
\usepackage{natbib}
\usepackage[dvips]{graphicx}
\usepackage[top=1in, bottom=1.20in, left=1.25in, right=1.0in]{geometry}
\usepackage[]{todonotes}
\usepackage{xcolor}

\usepackage{graphicx}
\usepackage{dcolumn}
\usepackage{bm}

\begin{document}


\title{Remote quantum clock synchronization without synchronized clocks}

\author{Ebubechukwu O. Ilo-Okeke}
\affiliation{New York University Shanghai, 1555 Century Ave, Pudong, Shanghai 200122, China}  

\author{Louis Tessler}
\affiliation{New York University Shanghai, 1555 Century Ave, Pudong, Shanghai 200122, China}  

\author{Jonathan P. Dowling}
\affiliation{Hearne Institute for Theoretical Physics, Department of Physics \& Astronomy, Louisiana State University, Baton Rouge, Louisiana 70803-4001, USA}

\author{Tim Byrnes}
\affiliation{New York University Shanghai, 1555 Century Ave, Pudong, Shanghai 200122, China}
\affiliation{Department of Physics, New York University, New York, NY 10003, USA}

\date{\today}

\begin{abstract}
A major outstanding problem for many quantum clock synchronization protocols is the hidden assumption of the availability of synchronized clocks within the protocol.  In general, quantum operations between two parties do not have consistent phase definitions of quantum states, which introduce an unknown systematic phase error.  We show that despite prior arguments to the contrary, it is possible to remove this unknown phase via entanglement purification.
This closes the loophole for entanglement based quantum clock synchronization protocols, which are most compatible with current photon based long-distance entanglement distribution schemes.  Starting with noisy Bell pairs, we show that the scheme produces a singlet state for any combination of (i) differing basis conventions for Alice and Bob; (ii) an overall time offset in the execution of the purification algorithm; and (iii) the presence of a noisy channel.  Error estimates reveal that better performance than existing classical Einstein synchronization protocols should be achievable using current technology.  
\end{abstract}

\maketitle


Access to a universally agreed global standard time is of great importance to many technologies such as data transfer networks, financial trading, airport traffic control, rail transportation networks, telecommunication networks, the global positioning system (GPS) and long baseline interferometry \cite{sundararaman2005clock}. To achieve this, clock synchronization is a fundamental task such that a network of clocks can be established, from which one can locally interrogate to obtain a common reference time. Classically, when special relativity is taken into account, there are two basic methods to synchronize clocks: Einstein synchronization \cite{einstein1905} and Eddington's slow clock transport \cite{eddington1924}.  In view of the superb stabilities that the next generation of atomic clocks are achieving \cite{ludlow2015optical}, the question of how best to synchronize clocks with high precision is one that must be addressed.  To address this demand, methods based on both ideas have been proposed for the accurate synchronization of clocks: time transfer laser links for the Einstein protocol \cite{samain2008time,samain2015time,giovannett2001,quan2016}, and quantum adaptations of Eddington's protocol \cite{chuang2000,zhang2004,burgh2005,tavakoli2015}.  A third method of clock synchronization, based on quantum entanglement, was proposed by Jozsa and co-workers which is independent of the relative locations or properties of the intervening medium  \cite{jozsa2000}. It uses shared prior entanglement between two clocks located at different spatial locations for synchronization. The original two party synchronization protocol \cite{jozsa2000,yurtsever2002,burt2001,jozsa2001} has been extended to multi-parties \cite{krco2002,benav2011,ren2012,komar2014}. Several experimental verifications of the protocol have been reported \cite{valencia2004,quan2016,zhang2004,kong2017}.   

\begin{figure}[t]
	\includegraphics[width=\columnwidth]{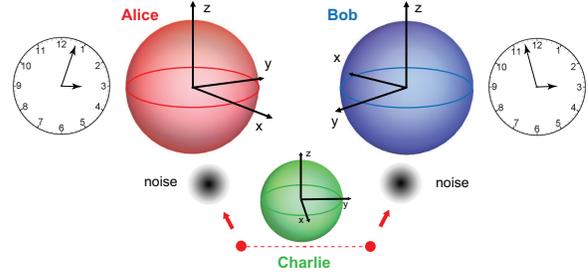}
	\caption{The situation considered for asynchronous quantum clock synchronization. Charlie distributes entangled singlet states to Alice and Bob, in his basis convention.  The entangled states are susceptible to noise, and become mixed on arrival at Alice and Bob's locations.  Alice and Bob have unsynchronized clocks, and also have different basis conventions for the coherent superpositions of the logical states $ |0 \rangle $ and $ | 1 \rangle $. By purifying many entangled qubits, the aim is to synchronize Alice and Bob's clocks. }
	\label{fig1}
\end{figure}

One major outstanding issue with many quantum clock synchronization (QCS) protocols is that they implicitly assume a common time reference \cite{preskill2000,yurtsever2002,burgh2005}.  The origin of this problem is that without the availability of synchronized clocks between Alice and Bob, definitions of superposition states of qubits such as $ (| 0 \rangle + | 1 \rangle )/\sqrt{2} $ are defined only up to a phase convention that is defined locally.  Worse still, any quantum algorithm that Alice and Bob execute may require careful synchronization in order to not introduce additional phases due to precession of the qubits.  This problem affects both quantum versions of Eddington and entanglement based schemes \cite{burgh2005,preskill2000,yurtsever2002}.  Proposals to overcome this issue have been proposed for Eddington schemes have been introduced, which require a two-way exchange of clock qubits \cite{burgh2005}.  This however involves sending clock qubit atoms (e.g. Cs, Rb, Sr) between the two parties, which is highly challenging for long-distance intercontinental or space-based communications.  In view of photonic long-distance space-based entanglement distribution now being demonstrated \cite{yin2017,ren2017ground}, a protocol compatible with this technology is most desirable. For example, long-distance entanglement could be first generated using photons, then stored on qubits where the clocks are present, then the QCS protocol of Ref. \cite{jozsa2000} can be executed.  We henceforth refer to the scheme of Ref. \cite{jozsa2000} when discussing ``QCS''.  

We show in this paper, contrary to previous arguments \cite{preskill2000}, that it is possible to produce an entangled state with controlled phase without Alice and Bob having any knowledge of each other's clocks.  The main observation is that the full distillation protocol as originally given by Bennett and co-workers \citep{bennett1996,bennett1996b} including random bilateral rotations ensures that the singlet state is produced, with respect to the {\it local basis choice}.  
Once this is prepared, it is possible to execute the original QCS protocol of Ref. \cite{jozsa2000}, despite the presence of additional phases, differing basis conventions, and noise. We assume that Alice and Bob do have clocks ticking at the correct frequency, such that they can keep track of the precession for the duration of the algorithm, but the clocks have in general a relative time offset (the clocks are syntonized but not synchronized) \cite{burt2001comment}.  The combination of the entanglement purification and the QCS allows for a completely asynchronous synchronization protocol for clocks, completing the scheme of Ref. \cite{jozsa2000}.

Suppose the singlet state 
\begin{align}
|\psi^- \rangle^{(C)}   =\frac{|1 \rangle_{A}^{(C)} | 0 \rangle_{B}^{(C)}   - |0 \rangle_{A}^{(C)}  | 1\rangle_{B}^{(C)}   }{\sqrt{2}} 
\label{idealsinglet}
\end{align}
is prepared and sent by Charlie to Alice and Bob.  Here the definitions of the states are with respect to Charlie's basis convention, which may be different to Alice and Bob's. Thus the state $ | 0 \rangle_{A}^{(C)} $ means a qubit state in Alice's possession, in the basis convention of Charlie, and so on.  We assume that Alice, Bob, and Charlie all have different basis conventions, which we can relate according to $| \sigma \rangle^{(A)}  = e^{-i \theta_{\sigma}^{(A)} } | \sigma \rangle^{(C)}$, $| \sigma \rangle^{(B)} = e^{-i \theta_{\sigma}^{(B)} } | \sigma \rangle^{(C)}$,
%
%
where $ \sigma \in \{ 0,1\} $. If the bases are transformed consistently using the same convention globally, then the state (\ref{idealsinglet}) is invariant, for example
\begin{align}
|\psi^- \rangle^{(B)}   = \frac{|1 \rangle_{A}^{(B)} | 0 \rangle_{B}^{(B)}   - |0 \rangle_{A}^{(B)}  | 1\rangle_{B}^{(B)}   }{\sqrt{2}},
\end{align}
where we chose the irrelevant global phase $ \theta_0^{(B)} + \theta_1^{{(B)}} = 0 $ for simplicity. However, as pointed out by Ref. \cite{preskill2000}, without the availability of synchronized clocks, it is not possible for Alice and Bob to know about their mutual basis conventions.  Thus the appropriate basis to view the state is in Alice and Bob's respective local bases
\begin{align}
& |\psi^- \rangle^{(\text{loc})} =  \nonumber \\
& \frac{ |1 \rangle_{A}^{(A)} | 0 \rangle_{B}^{(B)} - e^{i (\theta_{0}^{(A)} + \theta_{1}^{(B)} -\theta_{1}^{(A)} -\theta_{0}^{(B)} ) } |0 \rangle_{A}^{(A)}  | 1\rangle_{B}^{(B)} }{\sqrt{2}}.
\end{align}
We emphasize that $|\psi^- \rangle^{(\text{loc})} = |\psi^- \rangle^{(B)}  = |\psi^- \rangle^{(C)} $ are all in fact the same state, but they appear different due to different conventions. The effect of Alice and Bob choosing different basis conventions is equivalent to having an unknown relative phase in the singlet \cite{preskill2000,burgh2005}. We may define the relative difference between the basis choices of Alice and Bob by defining a rotation operator $U^{(AB)} | \sigma \rangle^{(B)} = | \sigma \rangle^{(A)}$, $U^{(BA)} | \sigma \rangle^{(A)}  = | \sigma \rangle^{(B)} $
%
%
which in this case is $U^{(AB)} = {U^{(BA)}}^\dagger = e^{i \sum_\sigma ( \theta_\sigma^{(B)} - \theta_\sigma^{(A)})|\sigma \rangle \langle \sigma |}$.
%
%
Operators then transform as 
\begin{align}
O^{(A)} = U^{(AB)} O^{(B)} {U^{(AB)}}^\dagger
\label{operatortrans}
\end{align}
and similarly for Bob's operators. 

In addition to the different basis conventions, when Alice and Bob perform their entanglement purification circuit, they will not know precisely when the other starts their first quantum operation.  Due to the precession of the qubits, there will be an additional phase offset in the singlet state, which without loss of generality we can attribute to Alice's side.  Hence the arriving singlet will have a form in the local basis (up to a global phase)
\begin{align}
\label{eq:p11}
\left|\psi^-_\varphi\right>^{(\text{loc})} & = T | \psi^- \rangle^{(\text{loc})} \nonumber \\
& =  \frac{1}{\sqrt{2}} ( |1 \rangle_{A}^{(A)} | 0 \rangle_{B}^{(B)}  - e^{i \varphi } |0 \rangle_{A}^{(A)}  | 1\rangle_{B}^{(B)} ), 
\end{align}
where the time delay operator is
\begin{align}
T = e^{-i\omega \delta t |1\rangle \langle 1 |},
\end{align}
$ \varphi = \theta_{0}^{(A)} + \theta_{1}^{(B)} -\theta_{1}^{(A)} - \theta_{0}^{(B)} - \omega \delta t $, and $ \delta t $ is the time difference between Alice and Bob's first quantum operation.  

Furthermore, in addition to the systematic error introduced by the different basis conventions and time offset, there may be a stochastic error which reduces the purity of the state.  We model this process using the noisy channel with both bit and phase flips, which for our state will appear as
\begin{align}
\label{eq:02}
\left|\psi^-_\varphi\right>^{(\text{loc})} & \left<\psi^-_\varphi\right|^{(\text{loc})}  \rightarrow \rho_\varphi^{(\text{loc})} \nonumber \\
& = \frac{p}{4} I + (1 - p ) \left|\psi^-_\varphi\right>^{(\text{loc})} \left<\psi^-_\varphi\right|^{(\text{loc})},
\end{align}
where $p$ is the probability that error will be introduced on sending the qubit through a noisy channel to Alice and Bob, and $ I $ is the $ 4 \times 4 $ identity matrix. We assume that $ N $ imperfect Bell pairs (\ref{eq:02}) are shared between Alice and Bob, and $ \varphi $ is unknown to both of them.  The task is then to achieve clock synchronization by first purifying the above state to a sufficiently high fidelity, then executing the QCS protocol without knowledge of any shared timing information.

\begin{figure}
\includegraphics[width=\columnwidth]{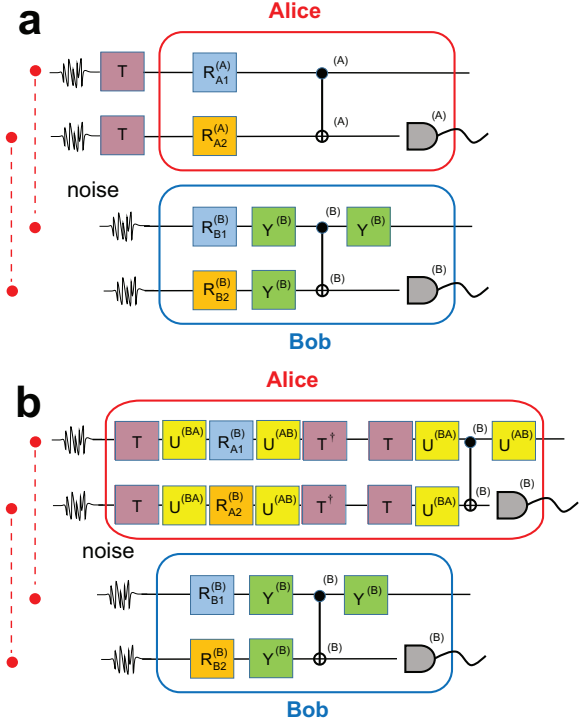}
		\caption{The quantum circuit for entanglement purification. Due to the lack of synchronized clocks between
		Alice and Bob, the basis choice for the circuit elements by each party will be in their respective bases, labeled by $ (A,B) $.  (a) The circuit as performed by Alice and Bob; (b) an equivalent circuit where all circuit elements have been transformed to the same basis choice.  $ {\cal B} = R_A \otimes R_B $ are random bilateral rotations which are predecided by Alice and Bob, $ U^{AB} $ transforms from Bob's basis convention to Alice's, $ T $ includes the effect of a time delay between the start of Alice and Bob's operations.  }
		\label{fig2}
\end{figure}

We first argue, using quantum circuit methods, that it is possible to perform entanglement purification such that a singlet state is obtained in the local basis. In the originally conceived form of entanglement purification \cite{bennett1996,bennett1996b}, the bilateral random unitary rotations and the Bell state comparison are performed synchronously, and also using the same basis convention throughout. In the context of QCS this cannot be performed, and instead the modified quantum circuit as shown in Fig. \ref{fig2}(a) will be executed. The noisy Bell states arriving will have a phase offset $ T $ which takes into account of any time delay between Alice and Bob's first operations.  Alice and Bob will furthermore execute the algorithm in their {\it local} basis conventions. We now deduce the effect of this circuit. We can rewrite the circuit in a form closer to the original, by applying the basis rotation (\ref{operatortrans}) around all the operators, and adding $ T^\dagger T = I $, which gives Fig. \ref{fig2}(b). After the first five operations of Alice in Fig. \ref{fig2}(b), and working in the common basis of Bob, Alice's random unitaries  are  transformed as $ T^\dagger {U^{(BA)}}^\dagger  R_A^{(B)} U^{(BA)} T $. This is the same as the standard bilateral rotations except that Alice's operations are transformed to a different basis. 
The state that results at this point of the circuit is the Werner state
\begin{align}
\rho_{\text W} & =  F |\psi^-_\varphi \rangle \langle \psi^-_\varphi | + \frac{1 - F}{3} ( I - |\psi^-_\varphi \rangle \langle \psi^-_\varphi | ).
\label{eq:09}
\end{align}
where $| \psi^-_\varphi \rangle  = T^\dagger {U^{(BA)}}^\dagger | \psi^- \rangle^{(B)} $.  At this point the states have an extra phase offset, but immediately after the bilateral rotation, the circuit operates with $ U^{(BA)} T $, which exactly cancels this extra factor (note the opposite operator ordering conventions for quantum circuits and equation form).  At this point we have a Werner state in Bob's basis, with various circuit elements all in Bob's basis.  The purification thus proceeds as originally conceived, purifying towards the state $  | \psi^- \rangle^{(B)}  $.  Finally, there is one extra  $ U^{(AB)}_A $ at the end of the circuit which completes the whole procedure.   The state that the purification thus converges to is thus
\begin{align}
\left|\psi^-_{\varphi=0}\right>^{(\text{loc})} &  = U^{(AB)}_A | \psi^- \rangle^{(B)} \nonumber \\
& = \frac{|1 \rangle_{A}^{(A)} | 0 \rangle_{B}^{(B)}   - |0 \rangle_{A}^{(A)}  | 1\rangle_{B}^{(B)}   }{\sqrt{2}} .
\label{localbasissinglet}
\end{align}
Thus starting from the state with an extra phase (\ref{eq:02}), the entanglement purification has converged to the singlet state with respect to the local basis choice with no relative phase $\varphi = 0$. The state (\ref{localbasissinglet}) is exactly as desired, since any measurements that will be made on this state will be in the local basis choice.  Alice and Bob would then measure this state in their local basis choice, which does not contain any extra phases as discussed in Ref. \cite{preskill2000}.  The QCS protocol then proceeds as described in Ref. \citep{jozsa2000}. 


This result can be also calculated by direct application of the bilateral rotations as shown in the Appendix.  Starting from the state $ \rho = p I/4 + (1-p)| \psi^-_\varphi \rangle^{(\text{loc})} \langle \psi^-_\varphi |^{(\text{loc})}$, we explicitly calculate that the bilateral rotations produce the state
\begin{align}
& \rho_{\text W}'  = I \frac{p}{4} + (1 -p ) \left|\psi^-_{\varphi = 0}\right>^{(\text{loc})} \left<\psi^-_{\varphi = 0}\right|^{(\text{loc})} \cos^2\left(\frac{\varphi}{2}\right) \nonumber \\
&  + \frac{1 - p}{3} \left[I - \left|\psi^-_{\varphi = 0}\right>^{(\text{loc})} \left<\psi^-_{\varphi = 0}\right|^{(\text{loc})} \right]\sin^2\left(\frac{\varphi}{2}\right) . \label{eq:08}
\end{align}
The fidelity calculated using (\ref{eq:08}) agrees with that calculated from (\ref{eq:02}) which gives $F = \langle \psi^- | \rho_\varphi |\psi^-\rangle = \frac{p}{4} + (1 - p)\cos^2\left(\frac{\varphi}{2}\right)$ since singlet states are invariant under bilateral rotations.The fidelity $ F $ contains an extra phase factor originating from the combination of the time delay and the different basis conventions.   Since (\ref{eq:08}) is in the local basis as desired, the remaining part of the purification proceeds in the regular way.  

\begin{figure}
\includegraphics[width=\columnwidth]{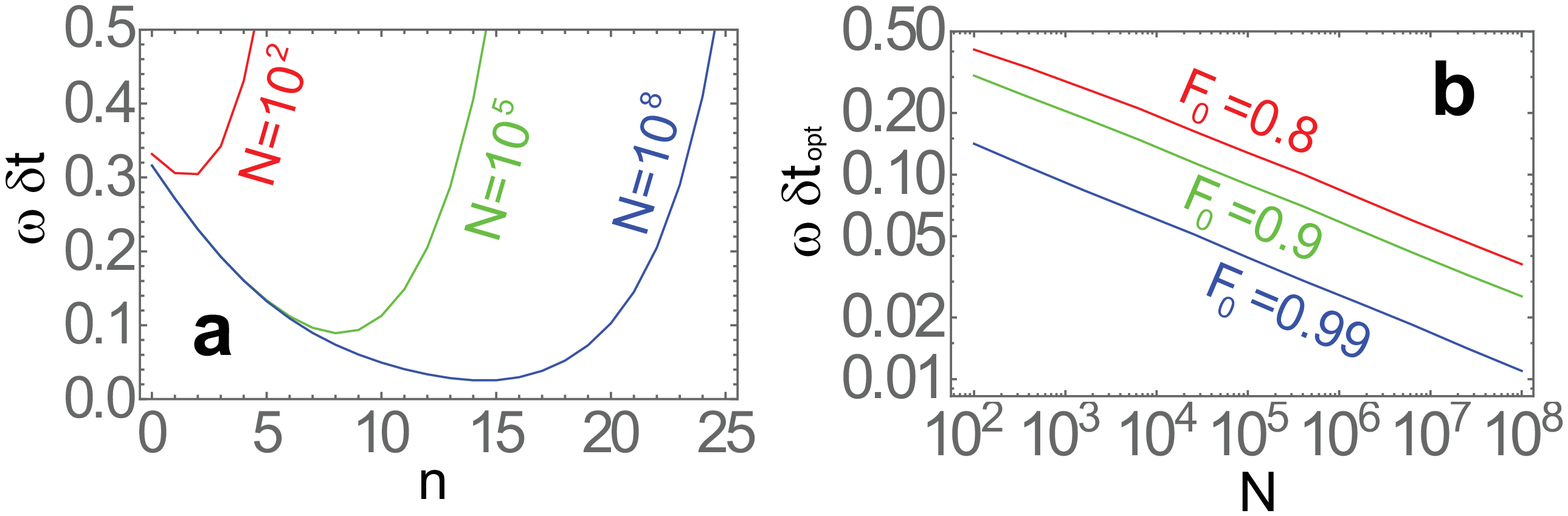}
		\caption{Accuracy of the QCS protocol. (a) The error in the QCS for  various numbers of available entangled Bell pairs $N$ as marked and $ F_0 = 0.9 $, as a function of purification rounds $ n $.  (b) Optimized error versus $ N $ for various initial fidelities $ F_0 $ as marked.  }
		\label{fig3}
\end{figure}

The above result removes an outstanding issue of the QCS protocol. What can we expect from a future implementation of QCS? To answer this we estimate the competitiveness of the QCS protocol in comparison to existing schemes.   Currently the most accurate long-distance clock synchronization protocols are microwave-based GPS and Two-way Satellite Time and Frequency Transfer (TWSTFT) \cite{allan1980accurate,kirchner1999two}, which achieve synchronizations at the level of 1 ns.  The next generation laser based methods aim to improve this to the level of 100 ps \cite{samain2008time,samain2015time}. The fundamental sources of error in the QCS protocol will be due to the imperfect entanglement that is distributed between Alice and Bob, and the quantum noise due to the standard quantum limit in the QCS protocol itself (see Appendix).  We estimate that the error in the QCS obeys a relation
\begin{align}
\delta t = \frac{1}{\omega} \sqrt{ \frac{2^n}{N} + 1 -F_n}
\end{align}
where $ \omega $ is the clock frequency, $ N $ is the number of available Bell pairs for QCS, $ n $ is the number of rounds of purification performed, and $ F_n $ is the fidelity of the Bell pairs after $ n $ rounds of purification.  In Fig. \ref{fig3}(a) we see that there is an optimum number of purification rounds.  This occurs because there is a trade-off between improving the fidelity of the Bell pairs by purification, and consuming Bell pairs for purification.  Using this optimum number of purification rounds, we obtain the level of accuracy expected in the QCS algorithm in Fig. \ref{fig3}(b).  As expected one obtains an improvement in performance with both $ N $ and $ F_0 $.  Taking currently achievable estimates for parameters  we have $ F_0 \approx 0.9 $ and $ N = 10^5 $ \cite{yin2017,ren2017ground}, and using the Cs clock transition frequency the timescale is set by $ \omega_{\text{Cs}}^{-1} = 17 $ ps, from which we obtain $ \delta t \approx 2 $ ps, a considerable improvement over classical schemes. Naturally, using larger numbers of Bell states and atoms with higher frequency clock transitions (e.g. Sr) one will obtain further improvements.

In summary, we have shown that using entanglement distillation it is possible for Alice and Bob to share a singlet state in their local basis, despite not having any information about their mutual basis conventions, and including any time offset between execution of their quantum gates.  The key ingredient is the incorporation of bilateral random unitaries in the entanglement purification protocol, which was not included in Ref. \cite{jozsa2000,preskill2000}.  This produces a Werner state in the local basis, and ``fixes'' the basis to a particular choice. This solves a major existing issue in the QCS protocol, where it was previously thought that synchronized clocks are required to perform the purification.  We have estimated the error of the protocol and found that it should have a performance that is considerably better than existing classical Einstein synchronization based schemes.  Here we only examined the same basic protocol as given in Ref. \cite{jozsa2000}, which has errors scaling as the standard quantum limit $ \propto 1/\sqrt{N} $.  Using collective states of the $ N $ Bell pairs should further improve the errors further to beat the standard quantum limit.  We envision that the QCS would be particularly useful in the context of the space-based quantum network \cite{yin2017,ren2017ground,byrnes2017}, where satellites are each in possession of an high-precision clock. Such entanglement based schemes are a powerful way to synchronize clocks without the use of a classical channel containing the timing information, which is susceptible to fluctuations in the atmosphere.  

The authors thank John Preskill for discussions. T. B. is supported by the Shanghai Research Challenge Fund; New York University Global Seed Grants for Collaborative Research; National Natural Science Foundation of China (Grant No. 61571301); the Thousand Talents Program for Distinguished Young Scholars (Grant No. D1210036A); and the NSFC Research Fund for International Young Scientists (Grant No. 11650110425); NYU-ECNU Institute of Physics at NYU Shanghai; and the Science and Technology Commission of Shanghai Municipality (Grant No. 17ZR1443600). J. P. D. would like to acknowledge support from the US Air Force Office of Scientific Research, the Army Research Office, the Defense Advanced Funding Agency, the National Science Foundation, and the Northrop-Grumman Corporation. E. O. I. O. acknowledges the  Talented Young Scientists Program (NGA-16-001) supported by the Ministry of Science and Technology of China.

\appendix
\section{Algebraic evaluation of the purification circuit} 

In this section we explicitly calculate the result of the purification circuit in Fig. 2(a), where there is an additional phase $ T $ due to the delay time and there is difference in basis choice of Alice and Bob. As given by (7), this effectively puts a phase offset in the singlet state in the local basis. The effect of random bilateral rotations \cite{bennett1996} is to put any state in the form of a Werner state as defined by the local Bell basis.  As discussed in Ref. \cite{bennett1996}, instead of applying an infinite set of random bilateral unitaries, it is equivalent to consider a finite set generated by $ {\cal G}_M = \sqrt{M_A^{(A)}} \otimes  \sqrt{M_B^{(B)}} $ with $ M \in \{X,Y,Z,I \} $.  
The first term in (7) in proportional to the identity, which is invariant under bilateral operations.  Since the identity is diagonal under any basis choice, we can choose equally the local basis 
\begin{align}
I = & |\psi^-_{\varphi = 0} \rangle^{(\text{loc})} \langle \psi^-_{\varphi = 0} |^{(\text{loc})} + |\psi^+_{\varphi = 0}\rangle^{(\text{loc})} \langle \psi^+_{\varphi = 0} |^{(\text{loc})} \nonumber \\
& + | \phi^-_{\varphi = 0}\rangle^{(\text{loc})} \langle \phi^-_{\varphi = 0} \rangle^{(\text{loc})}  + |\phi^+_{\varphi = 0}\rangle^{(\text{loc})} \langle \phi^+_{\varphi = 0} |^{(\text{loc})} .
\end{align}
For the second term in (7), we obtain by explicit computation
\begin{align}
& \sum_{n} {\cal B}_n | \psi^-_\varphi \rangle^{(\text{loc})} \langle \psi^-_\varphi |^{(\text{loc})} {\cal B}_n^\dagger  \nonumber \\
& =\frac{I - |\psi^-_{\varphi=0}\rangle^{(\text{loc})} \langle \psi^-_{\varphi=0} |^{(\text{loc})} }{3} \sin^2 (\frac{\varphi}{2} )  \nonumber \\
& +|\psi^-_{\varphi=0}\rangle^{(\text{loc})} \langle \psi^-_{\varphi=0} |^{(\text{loc})}  \cos^2 (\frac{\varphi}{2}),\label{eq:07}
\end{align}
where the sum is over the full group 
\begin{align}
{\cal B}_n \in &  \{ {\cal G}_I, {\cal G}_X {\cal G}_Y, {\cal G}_Y {\cal G}_Z, {\cal G}_Z {\cal G}_X, {\cal G}_X {\cal G}_Y {\cal G}_X {\cal G}_Y, \nonumber \\
&  {\cal G}_Y {\cal G}_Z {\cal G}_Y {\cal G}_Z, {\cal G}_Z {\cal G}_X {\cal G}_Z {\cal G}_X, {\cal G}_X {\cal G}_Z, \nonumber \\
& {\cal G}_X {\cal G}_Z {\cal G}_X {\cal G}_Z , {\cal G}_X {\cal G}_X, {\cal G}_Y {\cal G}_Y, {\cal G}_Z {\cal G}_Z \} ,
\end{align}
and is averaged over the number of group elements used in the rotation.
We compute the Werner state to be
\begin{align}
& \rho_{\text W}^\prime  = I \frac{p}{4} + (1 -p ) \left|\psi^-_{\varphi = 0}\right>^{(\text{loc})} \left<\psi^-_{\varphi = 0}\right|^{(\text{loc})} \cos^2\left(\frac{\varphi}{2}\right) \nonumber \\
&  + \frac{1 - p}{3} \left[I - \left|\psi^-_{\varphi = 0}\right>^{(\text{loc})} \left<\psi^-_{\varphi = 0}\right|^{(\text{loc})} \right]\sin^2\left(\frac{\varphi}{2}\right) , \label{eq:08}
\end{align}
as given in the main text.   The fidelity calculated using (\ref{eq:08}) agrees with that directly calculated from (7) which gives 
\begin{align}
F = \langle \psi^- | \rho_\varphi |\psi^-\rangle = \frac{p}{4} + (1 - p)\cos^2\left(\frac{\varphi}{2}\right)
\label{eq:06}
\end{align}
since singlet states are invariant under bilateral rotations.

This shows again that after the bilateral rotations, the state is correctly prepared in the local basis. Starting from a state (7) which had off-diagonal terms written in the local basis, the bilateral rotations have produced a state (\ref{eq:08}) that is diagonal.   With this preparation step, the remaining parts of the circuit in Fig. 2(a) can proceed in the normal way, since they are in the correct basis. We also see that the fidelity (\ref{eq:06}) contains an extra phase factor originating from the combination of the time delay and the different basis conventions.  This is natural since the phase $ \varphi $ will produce a state that is different from a singlet state, which will result in a loss of fidelity.  Since the purification protocol only works unless $ F>0.5 $, in practice this will mean that the phase will need to be controlled to some extent.  This can be achieved by having reasonably (but not exactly) synchronized clocks, so that the protocol can be executed accurately such that $ |\varphi | < \pi/2 $. For example, if QCS is performed periodically to counteract drift of the clocks, then by performing the protocol sufficiently frequently, the time offset can be bounded to within  $ |\varphi | < \pi/2 $.

\section{Error analysis of the quantum clock synchronization protocol}

Here we deduce the accuracy of the quantum clock synchronization (QCS) protocol in Ref. \cite{jozsa2000}.  The protocol proceeds as follows.  Alice and Bob prepares $ M $ Bell states (in the local basis, appropriate for use with the QCS protocol), after the necessary purification steps as given in the main text.  These states will in general have an imperfect fidelity with respect to ideal singlet states.  After the purification algorithm the state will be more realistically in a Werner state, but let us take the worst case scenario where the non-unit fidelity arises from a remnant phase error  $ \varepsilon $.  Such a phase error, which is a systematic error across all qubits, is the worst type of error for QCS as it gives an unknown time offset as we see below.  The state that is prepared prior to the QCS protocol is
\begin{align}
\prod_{n=1}^M   | \psi^-_{\varepsilon} \rangle^{(\text{loc})}_n   = & \prod_{n=1}^M \frac{1}{\sqrt{2}} \left( 
|1  \rangle_{nA}^{(A)} | 0\rangle_{nB}^{(B)} - e^{i \varepsilon} | 0 \rangle_{nA}^{(A)} | 1 \rangle_{nB}^{(B)}  \right)  \nonumber \\
= &\prod_{n=1}^M \frac{1}{2} \Big[ 
|+ \rangle_{nA}^{(A)} ( | 0\rangle_{nB}^{(B)} - e^{i \varepsilon} | 1 \rangle_{nB}^{(B)} ) \nonumber \\
& - |- \rangle_{nA}^{(A)} ( | 0\rangle_{nB}^{(B)} + e^{i \varepsilon} | 1 \rangle_{nB}^{(B)} )  \Big] ,
\label{initialstate0}
\end{align}
where the label $ nA $ and $ nB $ refers to the $ n $th qubit of Alice and Bob respectively.  We henceforth drop the labels $^{(A)}$, $^{(B)} $ and assume all operations are performed in the local basis.  Alice then performs a measurement in the $ | \pm \rangle $ basis, and tells Bob the measurement outcomes $ \sigma_n \in \{ 0,1 \} $.  After a time $ t $ from the measurement, Bob's qubit evolves to
\begin{align}
\prod_{n=1}^M \frac{1}{\sqrt{2}} \left( e^{-i\omega t/2} | 0 \rangle_{nB} + (-1)^{\sigma_n} e^{i(\omega t/2+\varepsilon)} | 1 \rangle_{nB} \right)
\end{align}
where precession of the qubits has occurred due to energy energy difference between the qubit states equal to $ \hbar \omega $.  When the data of Alice's measurement results arrives at Bob, Bob applies a $ Z $ gate to remove the $ (-1)^{\sigma_n} $ factor.  He then applies a Hadamard operation, which then gives the state
\begin{align}
\prod_{n=1}^M \left( \cos ( \frac{\omega t+\varepsilon}{2} )  | 0 \rangle_{nB}  - i \sin  ( \frac{\omega t+\varepsilon}{2} ) | 1 \rangle_{nB} \right) .
\end{align}
Bob then measures in the $ | 0 \rangle, | 1 \rangle $ basis, and obtains probabilities $ p_0 = \cos^2 (  (\omega t+\varepsilon_n)/2) $ and $ p_1 = \sin^2 (  (\omega t+\varepsilon_n)/2) $ for the two outcomes on a single qubit.  The probability of obtaining $ k $ qubits in the state $ | 0 \rangle $ and the remaining $ M - k $ in $ | 1 \rangle $ is 
\begin{align}
P_k & = { M \choose k } \cos^{2k} ( \frac{\omega t+\varepsilon}{2} ) \sin^{2k} ( \frac{\omega t+\varepsilon}{2} ) .
\end{align}
Using a Gaussian approximation \cite{ilookeke2014,Ilo-okeke2015} this can be written as
\begin{align}
P(x) \approx \frac{2}{| \sin \omega t |}\exp \left[ - \frac{M}{ \sin^2 ( \omega t +\varepsilon ) } \left( x - | \cos ( \omega t +\varepsilon ) | \right)^2 \right] .
\end{align}
where $ x  = \frac{2k-M}{M} $ is a probabilistic variable in the range $ [-1,1] $.  The time is then estimated according to this distribution which is sharply peaked at
\begin{align}
t = \frac{1}{\omega} \left( \cos^{-1} x - \varepsilon \right) .
\label{timeerror}
\end{align}
The variable $ x $ has a standard deviation
\begin{align}
\delta x  \approx \frac{1}{\sqrt{M}}
\end{align}
which then corresponds to an error in the time as 
\begin{align}
\delta t_{\text{SQL}} = \frac{1}{\omega \sqrt{M}} 
\end{align}
which scales as the standard quantum limit.  The phase error $ \varepsilon $ contributes also to an error in the time estimate (\ref{timeerror}).  This can be related to the fidelity by computing the overlap of (\ref{initialstate0}) with an ideal singlet state, which gives
\begin{align}
F & = | \langle  \psi^-_{\varepsilon=0} | \psi^-_{\varepsilon} \rangle |^2 = \cos^2 \frac{\varepsilon}{2} \nonumber \\
& \approx 1- \varepsilon^2.
\end{align}
The time error due to imperfect Bell pairs thus affects the QCS protocol according to 
\begin{align}
\delta t_{F} = \frac{\sqrt{1-F}}{\omega} .
\end{align}
The total error on the QCS protocol is then 
\begin{align}
\delta t & = \sqrt{ \delta^2 t_{\text{SQL}} + \delta^2 t_{F}} \nonumber \\
& = \frac{1}{\omega} \sqrt{ \frac{1}{M} + 1- F}
\end{align}
After $ n $ rounds of purification, the $ M $ Bell pairs are reduced to $ N = M/2^n $, and the fidelity is increased to \cite{bennett1996}
\begin{align}
F_{n} = \frac{F_{n-1}^2 + \frac{(1-F_{n-1})^2}{9}}{F_{n-1}^2 + \frac{2 F_{n-1} (1-F_{n-1})}{3} + \frac{5 (1-F_{n-1})^2}{9}},
\end{align}
which gives the result in the main text.


\begin{thebibliography}{33}
	\expandafter\ifx\csname natexlab\endcsname\relax\def\natexlab#1{#1}\fi
	\expandafter\ifx\csname bibnamefont\endcsname\relax
	\def\bibnamefont#1{#1}\fi
	\expandafter\ifx\csname bibfnamefont\endcsname\relax
	\def\bibfnamefont#1{#1}\fi
	\expandafter\ifx\csname citenamefont\endcsname\relax
	\def\citenamefont#1{#1}\fi
	\expandafter\ifx\csname url\endcsname\relax
	\def\url#1{\texttt{#1}}\fi
	\expandafter\ifx\csname urlprefix\endcsname\relax\def\urlprefix{URL }\fi
	\providecommand{\bibinfo}[2]{#2}
	\providecommand{\eprint}[2][]{\url{#2}}
	
	\bibitem[{\citenamefont{Sundararaman et~al.}(2005)\citenamefont{Sundararaman,
			Buy, and Kshemkalyani}}]{sundararaman2005clock}
	\bibinfo{author}{\bibfnamefont{B.}~\bibnamefont{Sundararaman}},
	\bibinfo{author}{\bibfnamefont{U.}~\bibnamefont{Buy}}, \bibnamefont{and}
	\bibinfo{author}{\bibfnamefont{A.~D.} \bibnamefont{Kshemkalyani}},
	\bibinfo{journal}{Ad hoc networks} \textbf{\bibinfo{volume}{3}},
	\bibinfo{pages}{281} (\bibinfo{year}{2005}).
	
	\bibitem[{\citenamefont{Einstein}(1905)}]{einstein1905}
	\bibinfo{author}{\bibfnamefont{A.}~\bibnamefont{Einstein}},
	\bibinfo{journal}{Annalen der Physik} \textbf{\bibinfo{volume}{17}},
	\bibinfo{pages}{891} (\bibinfo{year}{1905}).
	
	\bibitem[{\citenamefont{Eddington}(1924)}]{eddington1924}
	\bibinfo{author}{\bibfnamefont{A.~S.} \bibnamefont{Eddington}},
	\emph{\bibinfo{title}{The Mathematical Theory of Relativity}}
	(\bibinfo{publisher}{Cambridge University Press},
	\bibinfo{address}{Cambridge, England}, \bibinfo{year}{1924}).
	
	\bibitem[{\citenamefont{Ludlow et~al.}(2015)\citenamefont{Ludlow, Boyd, Ye,
			Peik, and Schmidt}}]{ludlow2015optical}
	\bibinfo{author}{\bibfnamefont{A.~D.} \bibnamefont{Ludlow}},
	\bibinfo{author}{\bibfnamefont{M.~M.} \bibnamefont{Boyd}},
	\bibinfo{author}{\bibfnamefont{J.}~\bibnamefont{Ye}},
	\bibinfo{author}{\bibfnamefont{E.}~\bibnamefont{Peik}}, \bibnamefont{and}
	\bibinfo{author}{\bibfnamefont{P.~O.} \bibnamefont{Schmidt}},
	\bibinfo{journal}{Reviews of Modern Physics} \textbf{\bibinfo{volume}{87}},
	\bibinfo{pages}{637} (\bibinfo{year}{2015}).
	
	\bibitem[{\citenamefont{Samain et~al.}(2008)\citenamefont{Samain, Weick,
			Vrancken, Para, Albanese, Paris, Torre, Zhao, Guillemot, and
			Petitbon}}]{samain2008time}
	\bibinfo{author}{\bibfnamefont{E.}~\bibnamefont{Samain}},
	\bibinfo{author}{\bibfnamefont{J.}~\bibnamefont{Weick}},
	\bibinfo{author}{\bibfnamefont{P.}~\bibnamefont{Vrancken}},
	\bibinfo{author}{\bibfnamefont{F.}~\bibnamefont{Para}},
	\bibinfo{author}{\bibfnamefont{D.}~\bibnamefont{Albanese}},
	\bibinfo{author}{\bibfnamefont{J.}~\bibnamefont{Paris}},
	\bibinfo{author}{\bibfnamefont{J.-M.} \bibnamefont{Torre}},
	\bibinfo{author}{\bibfnamefont{C.}~\bibnamefont{Zhao}},
	\bibinfo{author}{\bibfnamefont{P.}~\bibnamefont{Guillemot}},
	\bibnamefont{and} \bibinfo{author}{\bibfnamefont{I.}~\bibnamefont{Petitbon}},
	\bibinfo{journal}{International Journal of Modern Physics D}
	\textbf{\bibinfo{volume}{17}}, \bibinfo{pages}{1043} (\bibinfo{year}{2008}).
	
	\bibitem[{\citenamefont{Samain et~al.}(2015)\citenamefont{Samain, Exertier,
			Courde, Fridelance, Guillemot, Laas-Bourez, and Torre}}]{samain2015time}
	\bibinfo{author}{\bibfnamefont{E.}~\bibnamefont{Samain}},
	\bibinfo{author}{\bibfnamefont{P.}~\bibnamefont{Exertier}},
	\bibinfo{author}{\bibfnamefont{C.}~\bibnamefont{Courde}},
	\bibinfo{author}{\bibfnamefont{P.}~\bibnamefont{Fridelance}},
	\bibinfo{author}{\bibfnamefont{P.}~\bibnamefont{Guillemot}},
	\bibinfo{author}{\bibfnamefont{M.}~\bibnamefont{Laas-Bourez}},
	\bibnamefont{and} \bibinfo{author}{\bibfnamefont{J.~M.} \bibnamefont{Torre}},
	\bibinfo{journal}{Metrologia} \textbf{\bibinfo{volume}{52}},
	\bibinfo{pages}{423} (\bibinfo{year}{2015}).
	
	\bibitem[{\citenamefont{Giovannetti et~al.}(2001)\citenamefont{Giovannetti,
			Lloyd, and Maccone}}]{giovannett2001}
	\bibinfo{author}{\bibfnamefont{V.}~\bibnamefont{Giovannetti}},
	\bibinfo{author}{\bibfnamefont{S.}~\bibnamefont{Lloyd}}, \bibnamefont{and}
	\bibinfo{author}{\bibfnamefont{L.}~\bibnamefont{Maccone}},
	\bibinfo{journal}{Nature} \textbf{\bibinfo{volume}{412}},
	\bibinfo{pages}{417} (\bibinfo{year}{2001}).
	
	\bibitem[{\citenamefont{Quan et~al.}(2016)\citenamefont{Quan, Zhai, Wang, Hou,
			Wang, Xiang, Liu, Zhang, and Dong}}]{quan2016}
	\bibinfo{author}{\bibfnamefont{R.}~\bibnamefont{Quan}},
	\bibinfo{author}{\bibfnamefont{Y.}~\bibnamefont{Zhai}},
	\bibinfo{author}{\bibfnamefont{M.}~\bibnamefont{Wang}},
	\bibinfo{author}{\bibfnamefont{F.}~\bibnamefont{Hou}},
	\bibinfo{author}{\bibfnamefont{S.}~\bibnamefont{Wang}},
	\bibinfo{author}{\bibfnamefont{X.}~\bibnamefont{Xiang}},
	\bibinfo{author}{\bibfnamefont{T.}~\bibnamefont{Liu}},
	\bibinfo{author}{\bibfnamefont{S.}~\bibnamefont{Zhang}}, \bibnamefont{and}
	\bibinfo{author}{\bibfnamefont{R.}~\bibnamefont{Dong}},
	\bibinfo{journal}{Scientific Reports} \textbf{\bibinfo{volume}{6}},
	\bibinfo{pages}{30453} (\bibinfo{year}{2016}).
	
	\bibitem[{\citenamefont{Chuang}(2000)}]{chuang2000}
	\bibinfo{author}{\bibfnamefont{I.~L.} \bibnamefont{Chuang}},
	\bibinfo{journal}{Phys. Rev. Lett.} \textbf{\bibinfo{volume}{85}},
	\bibinfo{pages}{2006} (\bibinfo{year}{2000}).
	
	\bibitem[{\citenamefont{Zhang et~al.}(2004)\citenamefont{Zhang, Long, Deng,
			Liu, and Lu}}]{zhang2004}
	\bibinfo{author}{\bibfnamefont{J.}~\bibnamefont{Zhang}},
	\bibinfo{author}{\bibfnamefont{G.~L.} \bibnamefont{Long}},
	\bibinfo{author}{\bibfnamefont{Z.}~\bibnamefont{Deng}},
	\bibinfo{author}{\bibfnamefont{W.}~\bibnamefont{Liu}}, \bibnamefont{and}
	\bibinfo{author}{\bibfnamefont{Z.}~\bibnamefont{Lu}}, \bibinfo{journal}{Phys.
		Rev. A} \textbf{\bibinfo{volume}{70}}, \bibinfo{pages}{062322}
	(\bibinfo{year}{2004}).
	
	\bibitem[{\citenamefont{de~Burgh and Bartlett}(2005)}]{burgh2005}
	\bibinfo{author}{\bibfnamefont{M.}~\bibnamefont{de~Burgh}} \bibnamefont{and}
	\bibinfo{author}{\bibfnamefont{S.~D.} \bibnamefont{Bartlett}},
	\bibinfo{journal}{Phys. Rev. A} \textbf{\bibinfo{volume}{72}},
	\bibinfo{pages}{042301} (\bibinfo{year}{2005}).
	
	\bibitem[{\citenamefont{Tavakoli et~al.}(2015)\citenamefont{Tavakoli, Cabella,
			Zukowski, and Bourennane}}]{tavakoli2015}
	\bibinfo{author}{\bibfnamefont{A.}~\bibnamefont{Tavakoli}},
	\bibinfo{author}{\bibfnamefont{A.}~\bibnamefont{Cabella}},
	\bibinfo{author}{\bibfnamefont{M.}~\bibnamefont{Zukowski}}, \bibnamefont{and}
	\bibinfo{author}{\bibfnamefont{M.}~\bibnamefont{Bourennane}},
	\bibinfo{journal}{Scientific Reports} \textbf{\bibinfo{volume}{5}},
	\bibinfo{pages}{0782} (\bibinfo{year}{2015}).
	
	\bibitem[{\citenamefont{Jozsa et~al.}(2000)\citenamefont{Jozsa, Abrams,
			Dowling, and Williams}}]{jozsa2000}
	\bibinfo{author}{\bibfnamefont{R.}~\bibnamefont{Jozsa}},
	\bibinfo{author}{\bibfnamefont{D.~S.} \bibnamefont{Abrams}},
	\bibinfo{author}{\bibfnamefont{J.~P.} \bibnamefont{Dowling}},
	\bibnamefont{and} \bibinfo{author}{\bibfnamefont{C.~P.}
		\bibnamefont{Williams}}, \bibinfo{journal}{Phys. Rev. Lett.}
	\textbf{\bibinfo{volume}{85}}, \bibinfo{pages}{2010} (\bibinfo{year}{2000}).
	
	\bibitem[{\citenamefont{Yurtserver and Dowling}(2002)}]{yurtsever2002}
	\bibinfo{author}{\bibfnamefont{U.}~\bibnamefont{Yurtserver}} \bibnamefont{and}
	\bibinfo{author}{\bibfnamefont{J.~P.} \bibnamefont{Dowling}},
	\bibinfo{journal}{Phys. Rev. A} \textbf{\bibinfo{volume}{65}},
	\bibinfo{pages}{052317} (\bibinfo{year}{2002}).
	
	\bibitem[{\citenamefont{Burt et~al.}(2001{\natexlab{a}})\citenamefont{Burt,
			Ekstrom, and Swanson}}]{burt2001}
	\bibinfo{author}{\bibfnamefont{A.}~\bibnamefont{Burt}},
	\bibinfo{author}{\bibfnamefont{C.~R.} \bibnamefont{Ekstrom}},
	\bibnamefont{and} \bibinfo{author}{\bibfnamefont{T.~B.}
		\bibnamefont{Swanson}}, \bibinfo{journal}{Phys. Rev. Lett.}
	\textbf{\bibinfo{volume}{87}}, \bibinfo{pages}{129801}
	(\bibinfo{year}{2001}{\natexlab{a}}).
	
	\bibitem[{\citenamefont{Jozsa et~al.}(2001)\citenamefont{Jozsa, Abrams,
			Dowling, and Williams}}]{jozsa2001}
	\bibinfo{author}{\bibfnamefont{R.}~\bibnamefont{Jozsa}},
	\bibinfo{author}{\bibfnamefont{D.~S.} \bibnamefont{Abrams}},
	\bibinfo{author}{\bibfnamefont{J.~P.} \bibnamefont{Dowling}},
	\bibnamefont{and} \bibinfo{author}{\bibfnamefont{C.~P.}
		\bibnamefont{Williams}}, \bibinfo{journal}{Phys. Rev. Lett.}
	\textbf{\bibinfo{volume}{87}}, \bibinfo{pages}{129802}
	(\bibinfo{year}{2001}).
	
	\bibitem[{\citenamefont{Kr{\v c}o and Paul}(2002)}]{krco2002}
	\bibinfo{author}{\bibfnamefont{M.}~\bibnamefont{Kr{\v c}o}} \bibnamefont{and}
	\bibinfo{author}{\bibfnamefont{P.}~\bibnamefont{Paul}},
	\bibinfo{journal}{Phys. Rev. A} \textbf{\bibinfo{volume}{66}},
	\bibinfo{pages}{024305} (\bibinfo{year}{2002}).
	
	\bibitem[{\citenamefont{Ben-Av and Exman}(2011)}]{benav2011}
	\bibinfo{author}{\bibfnamefont{R.}~\bibnamefont{Ben-Av}} \bibnamefont{and}
	\bibinfo{author}{\bibfnamefont{I.}~\bibnamefont{Exman}},
	\bibinfo{journal}{Phys. Rev. A} \textbf{\bibinfo{volume}{84}},
	\bibinfo{pages}{014301} (\bibinfo{year}{2011}).
	
	\bibitem[{\citenamefont{Ren and Hofmann}(2012)}]{ren2012}
	\bibinfo{author}{\bibfnamefont{C.}~\bibnamefont{Ren}} \bibnamefont{and}
	\bibinfo{author}{\bibfnamefont{H.~F.} \bibnamefont{Hofmann}},
	\bibinfo{journal}{Phys. Rev. A} \textbf{\bibinfo{volume}{86}},
	\bibinfo{pages}{014301} (\bibinfo{year}{2012}).
	
	\bibitem[{\citenamefont{Komar et~al.}(2014)\citenamefont{Komar, Kessler,
			Bishof, Sorensen, Ye, and Lukin}}]{komar2014}
	\bibinfo{author}{\bibfnamefont{P.}~\bibnamefont{Komar}},
	\bibinfo{author}{\bibfnamefont{E.~M.} \bibnamefont{Kessler}},
	\bibinfo{author}{\bibfnamefont{L.}~\bibnamefont{Bishof},
		\bibfnamefont{M.;~Jiang}}, \bibinfo{author}{\bibfnamefont{A.~S.}
		\bibnamefont{Sorensen}},
	\bibinfo{author}{\bibfnamefont{J.}~\bibnamefont{Ye}}, \bibnamefont{and}
	\bibinfo{author}{\bibfnamefont{M.~D.} \bibnamefont{Lukin}},
	\bibinfo{journal}{Nature Physics} \textbf{\bibinfo{volume}{10}},
	\bibinfo{pages}{582} (\bibinfo{year}{2014}).
	
	\bibitem[{\citenamefont{Valencia et~al.}(2004)\citenamefont{Valencia,
			Scarcelli, and Shih}}]{valencia2004}
	\bibinfo{author}{\bibfnamefont{A.}~\bibnamefont{Valencia}},
	\bibinfo{author}{\bibfnamefont{G.}~\bibnamefont{Scarcelli}},
	\bibnamefont{and} \bibinfo{author}{\bibfnamefont{Y.}~\bibnamefont{Shih}},
	\bibinfo{journal}{Appl. Phys. Lett.} \textbf{\bibinfo{volume}{85}},
	\bibinfo{pages}{2655} (\bibinfo{year}{2004}).
	
	\bibitem[{\citenamefont{Kong et~al.}(2017)\citenamefont{Kong, Xin, Wei, Wang,
			Li, and Long}}]{kong2017}
	\bibinfo{author}{\bibfnamefont{X.}~\bibnamefont{Kong}},
	\bibinfo{author}{\bibfnamefont{T.}~\bibnamefont{Xin}},
	\bibinfo{author}{\bibfnamefont{S.}~\bibnamefont{Wei}},
	\bibinfo{author}{\bibfnamefont{B.}~\bibnamefont{Wang}},
	\bibinfo{author}{\bibfnamefont{K.}~\bibnamefont{Li}}, \bibnamefont{and}
	\bibinfo{author}{\bibfnamefont{G.}~\bibnamefont{Long}},
	\bibinfo{journal}{arXiv} \textbf{\bibinfo{volume}{quant-ph}},
	\bibinfo{pages}{1708.06050} (\bibinfo{year}{2017}).
	
	\bibitem[{\citenamefont{Preskill}(2000)}]{preskill2000}
	\bibinfo{author}{\bibfnamefont{J.}~\bibnamefont{Preskill}},
	\bibinfo{journal}{arXiv} \textbf{\bibinfo{volume}{quant-ph}},
	\bibinfo{pages}{0010098v1} (\bibinfo{year}{2000}).
	
	\bibitem[{\citenamefont{Yin et~al.}(2017)\citenamefont{Yin, Cao, Li, Liao,
			Zhang, Ren, Cai, Liu, Li, Dai et~al.}}]{yin2017}
	\bibinfo{author}{\bibfnamefont{J.}~\bibnamefont{Yin}},
	\bibinfo{author}{\bibfnamefont{Y.}~\bibnamefont{Cao}},
	\bibinfo{author}{\bibfnamefont{Y.~H.} \bibnamefont{Li}},
	\bibinfo{author}{\bibfnamefont{S.~K.} \bibnamefont{Liao}},
	\bibinfo{author}{\bibfnamefont{L.}~\bibnamefont{Zhang}},
	\bibinfo{author}{\bibfnamefont{J.~G.} \bibnamefont{Ren}},
	\bibinfo{author}{\bibfnamefont{W.~Q.} \bibnamefont{Cai}},
	\bibinfo{author}{\bibfnamefont{W.~Y.} \bibnamefont{Liu}},
	\bibinfo{author}{\bibfnamefont{B.}~\bibnamefont{Li}},
	\bibinfo{author}{\bibfnamefont{H.}~\bibnamefont{Dai}}, \bibnamefont{et~al.},
	\bibinfo{journal}{Science} \textbf{\bibinfo{volume}{356}},
	\bibinfo{pages}{1140} (\bibinfo{year}{2017}).
	
	\bibitem[{\citenamefont{Ren et~al.}(2017)\citenamefont{Ren, Xu, Yong, Zhang,
			Liao, Yin, Liu, Cai, Yang, Li et~al.}}]{ren2017ground}
	\bibinfo{author}{\bibfnamefont{J.-G.} \bibnamefont{Ren}},
	\bibinfo{author}{\bibfnamefont{P.}~\bibnamefont{Xu}},
	\bibinfo{author}{\bibfnamefont{H.-L.} \bibnamefont{Yong}},
	\bibinfo{author}{\bibfnamefont{L.}~\bibnamefont{Zhang}},
	\bibinfo{author}{\bibfnamefont{S.-K.} \bibnamefont{Liao}},
	\bibinfo{author}{\bibfnamefont{J.}~\bibnamefont{Yin}},
	\bibinfo{author}{\bibfnamefont{W.-Y.} \bibnamefont{Liu}},
	\bibinfo{author}{\bibfnamefont{W.-Q.} \bibnamefont{Cai}},
	\bibinfo{author}{\bibfnamefont{M.}~\bibnamefont{Yang}},
	\bibinfo{author}{\bibfnamefont{L.}~\bibnamefont{Li}}, \bibnamefont{et~al.},
	\bibinfo{journal}{arXiv preprint arXiv:1707.00934}  (\bibinfo{year}{2017}).
	
	\bibitem[{\citenamefont{Bennett
			et~al.}(1996{\natexlab{a}})\citenamefont{Bennett, Brassard, Popescu,
			Schumacher, Smolin, and Wootters}}]{bennett1996}
	\bibinfo{author}{\bibfnamefont{C.~H.} \bibnamefont{Bennett}},
	\bibinfo{author}{\bibfnamefont{G.}~\bibnamefont{Brassard}},
	\bibinfo{author}{\bibfnamefont{S.}~\bibnamefont{Popescu}},
	\bibinfo{author}{\bibfnamefont{B.}~\bibnamefont{Schumacher}},
	\bibinfo{author}{\bibfnamefont{J.~A.} \bibnamefont{Smolin}},
	\bibnamefont{and} \bibinfo{author}{\bibfnamefont{W.~K.}
		\bibnamefont{Wootters}}, \bibinfo{journal}{Phys. Rev. Lett.}
	\textbf{\bibinfo{volume}{76}}, \bibinfo{pages}{722}
	(\bibinfo{year}{1996}{\natexlab{a}}).
	
	\bibitem[{\citenamefont{Bennett
			et~al.}(1996{\natexlab{b}})\citenamefont{Bennett, {DiVincenzo}, Smolin, and
			Wootters}}]{bennett1996b}
	\bibinfo{author}{\bibfnamefont{C.~H.} \bibnamefont{Bennett}},
	\bibinfo{author}{\bibfnamefont{D.~P.} \bibnamefont{{DiVincenzo}}},
	\bibinfo{author}{\bibfnamefont{J.~A.} \bibnamefont{Smolin}},
	\bibnamefont{and} \bibinfo{author}{\bibfnamefont{W.~K.}
		\bibnamefont{Wootters}}, \bibinfo{journal}{Phys. Rev. A}
	\textbf{\bibinfo{volume}{54}}, \bibinfo{pages}{3824}
	(\bibinfo{year}{1996}{\natexlab{b}}).
	
	\bibitem[{\citenamefont{Burt et~al.}(2001{\natexlab{b}})\citenamefont{Burt,
			Ekstrom, and Swanson}}]{burt2001comment}
	\bibinfo{author}{\bibfnamefont{E.~A.} \bibnamefont{Burt}},
	\bibinfo{author}{\bibfnamefont{C.~R.} \bibnamefont{Ekstrom}},
	\bibnamefont{and} \bibinfo{author}{\bibfnamefont{T.~B.}
		\bibnamefont{Swanson}}, \bibinfo{journal}{Physical review letters}
	\textbf{\bibinfo{volume}{87}}, \bibinfo{pages}{129801}
	(\bibinfo{year}{2001}{\natexlab{b}}).
	
	\bibitem[{\citenamefont{Allan and Weiss}(1980)}]{allan1980accurate}
	\bibinfo{author}{\bibfnamefont{D.~W.} \bibnamefont{Allan}} \bibnamefont{and}
	\bibinfo{author}{\bibfnamefont{M.~A.} \bibnamefont{Weiss}}, in
	\emph{\bibinfo{booktitle}{34th Annual Symposium on Frequency Control. 1980}}
	(\bibinfo{organization}{IEEE}, \bibinfo{year}{1980}), pp.
	\bibinfo{pages}{334--346}.
	
	\bibitem[{\citenamefont{Kirchner}(1999)}]{kirchner1999two}
	\bibinfo{author}{\bibfnamefont{D.}~\bibnamefont{Kirchner}},
	\bibinfo{journal}{Review of Radio Science}  (\bibinfo{year}{1999}).
	
	\bibitem[{\citenamefont{Byrnes et~al.}(2017)\citenamefont{Byrnes, Ilyas,
			Tessler, Jambulingam, and Dowling}}]{byrnes2017}
	\bibinfo{author}{\bibfnamefont{T.}~\bibnamefont{Byrnes}},
	\bibinfo{author}{\bibfnamefont{B.}~\bibnamefont{Ilyas}},
	\bibinfo{author}{\bibfnamefont{L.}~\bibnamefont{Tessler}},
	\bibinfo{author}{\bibfnamefont{S.}~\bibnamefont{Jambulingam}},
	\bibnamefont{and} \bibinfo{author}{\bibfnamefont{J.~P.}
		\bibnamefont{Dowling}}, \bibinfo{journal}{arXiv}
	\textbf{\bibinfo{volume}{quant-ph}}, \bibinfo{pages}{1704.04774}
	(\bibinfo{year}{2017}).
	
	\bibitem[{\citenamefont{Ilo-Okeke and Byrnes}(2014)}]{ilookeke2014}
	\bibinfo{author}{\bibfnamefont{E.~O.} \bibnamefont{Ilo-Okeke}}
	\bibnamefont{and} \bibinfo{author}{\bibfnamefont{T.}~\bibnamefont{Byrnes}},
	\bibinfo{journal}{Phys. Rev. Lett.} \textbf{\bibinfo{volume}{112}},
	\bibinfo{pages}{233602} (\bibinfo{year}{2014}).
	
	\bibitem[{\citenamefont{Ilo-Okeke and Byrnes}(2016)}]{Ilo-okeke2015}
	\bibinfo{author}{\bibfnamefont{E.~O.} \bibnamefont{Ilo-Okeke}}
	\bibnamefont{and} \bibinfo{author}{\bibfnamefont{T.}~\bibnamefont{Byrnes}},
	\bibinfo{journal}{Phys. Rev. A} \textbf{\bibinfo{volume}{94}},
	\bibinfo{pages}{013617} (\bibinfo{year}{2016}).
	
\end{thebibliography}

%
%
\end{document}